\documentclass[%
reprint,
superscriptaddress,
 amsmath,amssymb,
 aps,
prb,
]{revtex4-2}

\usepackage{graphicx}
\usepackage{dcolumn}
\usepackage{bm}
\usepackage{siunitx}
\usepackage{xr}
\usepackage[version=4]{mhchem}
\usepackage[english]{babel}

\DeclareSIUnit{\PhiO}{\ensuremath{\Phi_{0}}}
\DeclareSIUnit{\electronvolt}{eV}
\DeclareSIUnit{\torr}{Torr}
\DeclareSIUnit{\atoms}{atoms}
\newcommand{\dMdV}{$\mathrm{d}M/\mathrm{d}V_{\mathrm{TG}}$}

\newcommand{\CrBST}{$\mathrm{Cr_{0.15}(Bi,Sb)_{1.85}Te_{3}}$}
\newcommand{\BST}{$\mathrm{(Bi,Sb)_{1.85}Te_{3}}$}
\newcommand{\CrST}{$\mathrm{Cr_{0.15}Sb_{1.85}Te_{3}}$}

\newcommand{\STO}{SrTiO$\mathrm{_3}$}

\begin{document}

\preprint{APS/123-QED}

\title{Imaging signatures of edge currents in a magnetic topological insulator}

\author{G. M. Ferguson}
\affiliation{
Department of Physics, Laboratory of Atomic and Solid State Physics,
Cornell University, Ithaca, New York 14853, USA
}
 \altaffiliation[Now at ]{Max Planck Institute for Chemical Physics of Solids}
\author{Run Xiao}
\affiliation{
Department of Physics, The Pennsylvania State University, University Park, 16802, Pennsylvania, USA
}
\author{Anthony R. Richardella}
\affiliation{
Department of Physics, The Pennsylvania State University, University Park, 16802, Pennsylvania, USA
}
\author{Austin Kaczmarek}
\affiliation{
Department of Physics, Laboratory of Atomic and Solid State Physics,
Cornell University, Ithaca, New York 14853, USA
}
\author{Nitin Samarth}
\affiliation{
Department of Physics, The Pennsylvania State University, University Park, 16802, Pennsylvania, USA
}
\author{Katja C. Nowack}%
 \email{kcn34@cornell.edu}
 \affiliation{
Department of Physics, Laboratory of Atomic and Solid State Physics,
Cornell University, Ithaca, New York 14853, USA
}
\affiliation{
 Kavli Institute at Cornell for Nanoscale Science, Ithaca, New York 14853, USA
}


\begin{abstract}

Magnetic topological insulators (MTIs) host topologically protected edge states, but the role that these edge states play in electronic transport remains unclear. Using scanning superconducting quantum interference device (SQUID) microscopy, we performed local measurements of the current distribution in a quantum anomalous Hall (QAH) insulator at large bias currents, where the quantization of the conductivity tensor breaks down. We find that bulk currents in the channel interior coexist with edge currents at the sample boundary. While the position of the edge current changes with the reversal of the magnetic field, it does not depend on the current direction. To understand our observations, we introduce a model which includes contributions from both the sample magnetization and currents driven by chemical potential gradients. To parameterize our model, we use local measurements of the chemical potential induced changes in the sample magnetization. Our model reveals that the observed edge currents can be understood as changes in the magnetization generated by the electrochemical potential distribution in the sample under bias. Our work underscores the complexity of electronic transport in MTIs and highlights both the value and challenges of using magnetic imaging to disentangle various contributions to the electronic transport signatures.


\end{abstract}

\maketitle
Magnetic topological insulators (MTIs) exhibit a wealth of unusual and potentially useful electronic transport phenomena. Most prominently, the quantum anomalous Hall effect (QAHE), characterized by a quantized Hall conductivity and dissipationless electronic transport, has been observed in a growing list of MTIs \cite{chang2013experimental, deng2020quantum, serlin2020intrinsic, li2021quantum, han2024correlated, han2024large, choi2024electric, chang2023review}. MTIs have also demonstrated large electronically switchable Hall effects \cite{serlin2020intrinsic, yuan2024electrical}, large non-reciprocal transport \cite{yasuda2020large}, and giant spin-orbit torques \cite{fan2014magnetization, fan2016electric, kondou2016fermi}. Expanding experimental control over these transport phenomena has been driven by interest in adapting MTIs for a range of different applications in metrology \cite{okazaki2022quantum, rodenbach2022metrological}, quantum information processing \cite{lian2018topological}, and spintronics \cite{fan2014magnetization, fan2016electric, yasuda2017current, yuan2024electrical}.


The phenomenology of MTIs is often attributed to topologically protected chiral edge states, which are predicted to be present at the sample boundary. A source-drain bias may shift the chemical potential of the edge states, causing a non-equilibrium current to flow along the sample edge \cite{buttiker1988absence}. Similarly, a uniform shift in the chemical potential of a MTI generates a shift in the sample magnetization \cite{zhu2020voltage, tschirhart2021imaging}.

Previously, we used magnetic imaging to investigate the current distribution in \CrBST{} samples including the one reported here in the low-bias limit, where they exhibit the QAHE. We found that electronic transport was dominated by the bulk when the conductivity tensor was quantized \cite{ferguson2023direct}, indicating that bulk currents driven by the electric field in the sample interior play an important role in electronic transport inside the QAH regime \cite{weis2011metrology, rosen2022measured}. Here, we used scanning SQUID microscopy to investigate the current distribution at much larger bias currents, where $\rho_{xy}$ deviates from the quantized value and the sample exhibits a significant $\rho_{xx}$. 

We studied two Hall bars with identical dimensions, one fabricated from \CrBST{} and the other from \CrST{}. Both samples were grown on \ce{SrTiO3} substrates and feature a global back gate, which we used to tune the chemical potential in the channel, as well as a top gate. The first device, depicted schematically in Fig. 1a, is fabricated from \CrBST{} and exhibits the QAHE at low bias currents. The device fabricated from \CrST{}, exhibits only a small anomalous Hall effect as the Fermi level lies far below the gap. Further information on the electronic transport behavior of both devices as well as the current distribution in the low-bias limit was presented in Ref. \cite{ferguson2023direct} as Devices C and D, respectively. 

To determine the current distribution in the sample, we scanned a micrometer-scale SQUID \cite{huber2008gradiometric} approximately \SI{1}{\micro\meter} above the sample surface while sourcing a sinusoidal current with zero DC offset through the two large current contacts. The stray magnetic field generated by the current distribution in the sample coupled a flux $\Phi_{AC}$ into the pickup loop of the SQUID, which we detect using a lock-in amplifier. Further details of the sample fabrication and scanning SQUID measurements are described in Appendices A and B, respectively. 

\begin{figure}[h]
    \centering
    \includegraphics[width=0.45\textwidth]{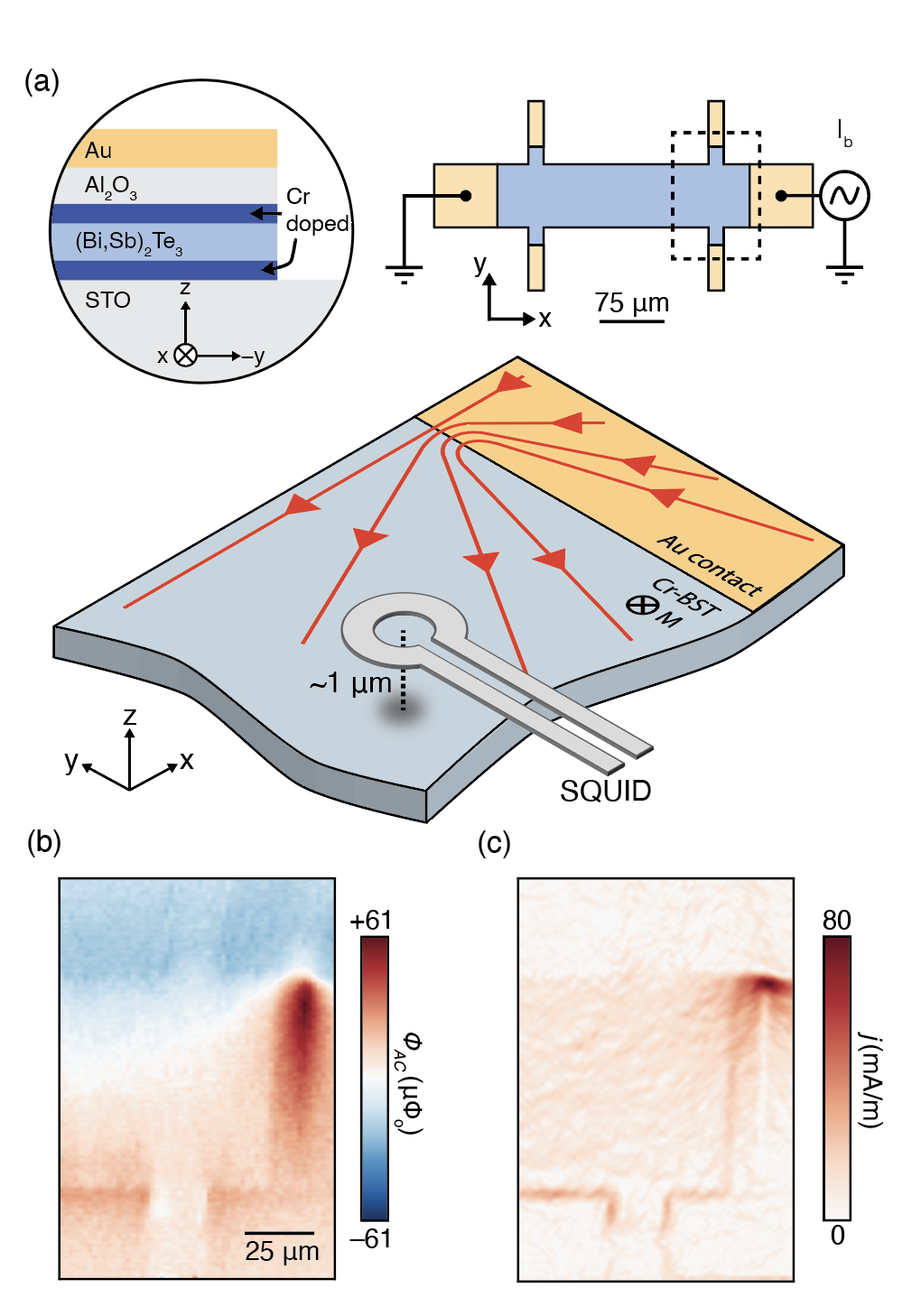}
    \caption{(a) Schematic depiction of the \CrBST{} device and scanning SQUID measurement. Left: cross-sectional view of the device. Two heavily Cr-doped layers of \BST{} sandwiching an undoped layer of \BST{} are grown on an \STO{} substrate. Gating through the \STO{} substrate is used to tune the chemical potential. Right: top-down view of the device. The topological insulator (blue) is patterned into a Hall-bar geometry with Au (yellow) contacts. The right contact is used to source a sinusoidal bias current $I_{b}$ through the channel while the left contact is grounded. Bottom: A Nb SQUID detects the stray magnetic field generated by the current distribution (red) in the channel. (b) Image of the flux coupled into the SQUID pickup loop in response to the sinusoidal bias current ($\Phi_{AC}$), acquired in the field of view indicated by the dashed line in (a). Here $I_{b} = \SI{500}{\nano\ampere}$ RMS, the sample is magnetized into the plane, and the sample is tuned into the magnetic exchange gap with $V_{BG} = \SI{110}{\volt}$. (c) Magnitude of the reconstructed  current density from the $\Phi_{AC}$ image in (b).}
    \label{fig:fig1}
\end{figure}

In Fig. 1b, we show an image of $\Phi_{AC}$ near the interface between the Au contact and the \CrBST{} channel, acquired with a \SI{500}{\nano\ampere} RMS bias current. Before acquiring the image, we magnetized the sample perpendicular to the device plane (along the $-\hat{z}$ direction), and tuned it into the magnetic exchange gap by applying a back gate voltage of $V_{BG} = \SI{110}{\volt}$ \cite{ferguson2023direct}. We acquired all of the magnetic imaging data reported here in zero applied magnetic field and with the sample cooled below \SI{50}{\milli\kelvin}. We observe a strong $\Phi_{AC}$ signal near the top corner of the contact which becomes progressively weaker away from the interface between the contact and the device channel. 

To determine the current distribution corresponding to the observed flux signal, we used a deconvolution approach based on Bayesian inference described in Ref. \cite{clement2019reconstruction} and in Appendix: Current Reconstruction. In Fig. 1c, we show the magnitude, $j$, of the resulting current density inside the sample, reconstructed from the $\Phi_{AC}$ image in Fig. 1b with $j = \sqrt{j_{x}^2 + j_{y}^2}$ and $j_x$ and $j_y$ the $x$ and $y$ components of the current density. We find that the strong magnetic flux signature at the top corner of the contact in the $\Phi_{AC}$ image results from the majority of the current entering the device through a narrow region at the interface between the Au contact and \CrBST{} channel. Away from the contact, the current spreads out over the width of the channel. In addition, we observe an enhanced current density traversing the width of the channel where the top gate of our sample ends, reaching the bottom edge, then flowing along the bottom edge of the device and appearing to enter and exit the voltage probe.

At the interface between the \CrBST{} channel and Au contact, the Hall angle $\theta_H$ between the current and the electric field given by $\tan \left(\theta_{H}\right) = \rho_{xy}/\rho_{xx}$ changes dramatically. In electrostatic potential simulations, this large change leads to the current density entering at one corner of the contact. Previous experiments on the IQHE \cite{klass1991fountain, klass1992image, komiyama2006electron, kawano1999cyclotron}, electrostatic potential simulations \cite{kirtley1986voltage, rosen2022measured} and our scanning SQUID measurements (See Supplementary Information Section: \ref{si_electrostatics}) confirm that location of the current injection is determined by the sign of $\rho_{xy}$. The details of the hot spot behavior and its associated heating effects will be addressed in a separate manuscript. In the following, we focus on understanding the observed enhanced current density along the device edges, which is not explained by the electrostatic potential simulations.

Within the limits of our experimental precision, all of the current is injected through the top corner of the contact. Away from the contact, we estimate that the enhanced current density along the edge or edge current corresponds to approximately 10\% of the total current flowing within \SI{3}{\micro\meter} of the sample edge. This is an upper bound on the confinement of the enhanced current density at the edge limited by the spatial resolution $\sim \SI{3}{\micro\meter}$ of our magnetic imaging technique \cite{ferguson2023direct}.

We first investigate the dependence of the edge current on the sample magnetization and $V_{BG}$. Fig. 2a shows the magnitude of the reconstructed current density between the voltage probes of the Hall bar, away from the contacts, with a source-drain current of \SI{2}{\micro\ampere} flowing through the channel. The sample magnetization and $V_{BG}$ are the same as in Fig. 1, and we again observe an increased current density along the bottom edge of the sample. Fig. 2b shows the same measurement with the sample magnetization reversed, revealing that when the magnetization is out of the plane, the top edge of the sample exhibits an enhanced current density. 

The magnitude of the edge current depends on $V_{BG}$. We found that tuning the chemical potential into the surface-state valence band with $V_{BG} = \SI{0}{\volt}$ significantly reduces the strength of the edge currents for both magnetization directions (Fig. 2c-d). Magnetic imaging of a second device, fabricated from \CrST{}, where the chemical potential lies deep in the valence band, showed no signatures of edge currents (see Supplementary Information Section \ref{si_CRST}).

\begin{figure}
    \centering
    \includegraphics[width=0.5\textwidth]{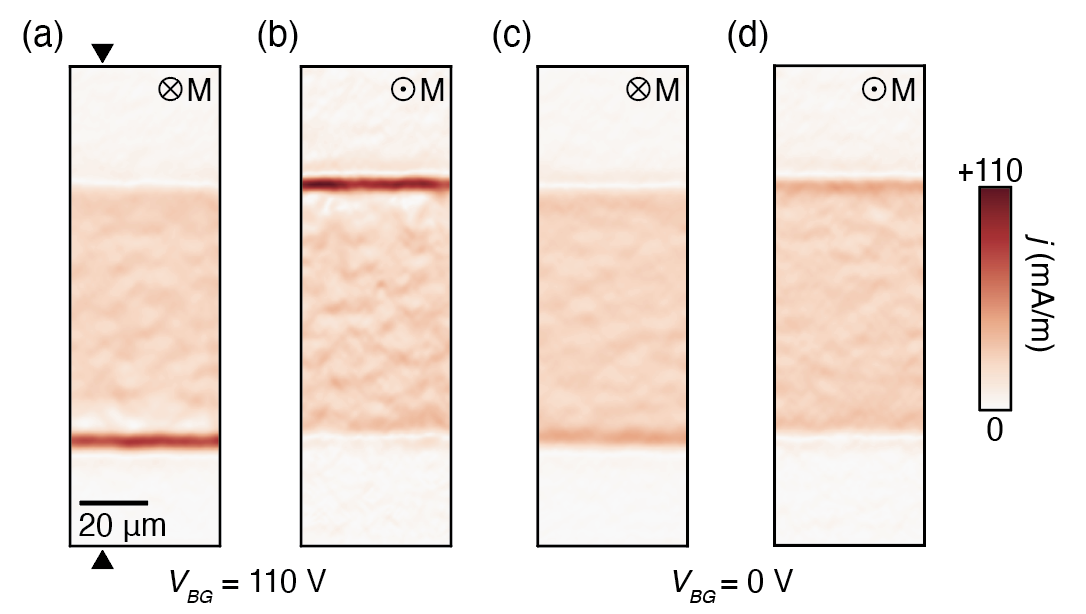}
    \caption{(a) Reconstructed current density between the voltage probes of the Hall bar with $I_{b} = \SI{2}{\micro\ampere}$, the sample magnetized into the plane, and the device tuned into the magnetic exchange gap with $V_{BG} = \SI{110}{\volt}$. An edge current is observed on the bottom edge of the device. (b) Same as (a), with the magnetization direction reversed. An edge current is observed on the top edge of the sample. (c) Same as (a), with the device tuned out of the magnetic exchange gap with $V_{BG} = \SI{0}{\volt}$. The strength of the edge current is dramatically reduced when the chemical potential lies outside the magnetic exchange gap. (d) Same as (c), with the magnetization reversed.}.
    \label{fig:fig2}
\end{figure}

To study the dependence of the current distribution on $V_{BG}$ in greater detail, we present line-cuts of the reconstructed current density as a function of $V_{BG}$ in Fig. 3a. Although we observed an enhanced current density for all back gate voltages along the bottom (top) edge of the channel when the sample is magnetized into (out of) the plane, the edge current signature is strongest in the neighborhood of $V_{BG} = \SI{110}{\volt}$, when the sample is gated into the magnetic exchange gap. 

While one edge of the sample shows an enhanced current density, the current density at the opposite edge is slightly suppressed. In Figs. 3a and b, the suppression of the current density appears as an apparent shift in the portion of the sample which supports a downstream current density between the two magnetization directions. Although the top edge of the sample exhibits an enhanced current density when the sample is magnetized out-of the plane (Fig. 3a), when the magnetization is reversed (Fig. 3b), the current density in the same region of the sample is close to zero. Similarly, the portion of the sample at the bottom edge which supports an enhanced current density in Fig. 3b appears to carry no current once the magnetization is reversed. This effect can also be seen by closely comparing Figs. 2a and 2b. In Fig. 2b, current appears at the top edge where none is present in Fig. 2a. Figs. 3a and b show that this effect is most pronounced when $V_{BG}$ is chosen to tune the sample outside the magnetic exchange gap.

The current distributions that we observe in Fig. 1c, and Fig. 2 are inconsistent with helical or topologically trivial edge channels \cite{chang2015zero}. Previous magnetic imaging studies of the current distribution in both topologically protected helical edge states \cite{nowack2013imaging, spanton2014images}, and trivial edge states \cite{nichele2016edge} have shown that current flows along both edges of the sample in parallel. In contrast, our device shows a strong edge current only along one side of the channel, with the side determined by the sample magnetization. As shown in Fig. 1c, the current bypasses the top edge of the sample entirely even though it is injected in the top corner of the contact. In a helical or trivial model for the edge channels, the top edge of the sample would offer a path of similar resistance and would be expected to carry an edge current of similar magnitude to the bottom edge. The asymmetry in current density between the two edges is also inconsistent with edge transport models in which the current distribution shifts between the edges upon the reversal of the bias current. Our lock-in measurement scheme sources an AC current through the channel which averages over currents in the $+\hat{x}$ and $-\hat{x}$ directions. This would appear as both edges conducting in parallel in our measurements. 

\begin{figure}
    \centering
    \includegraphics[width=0.5\textwidth]{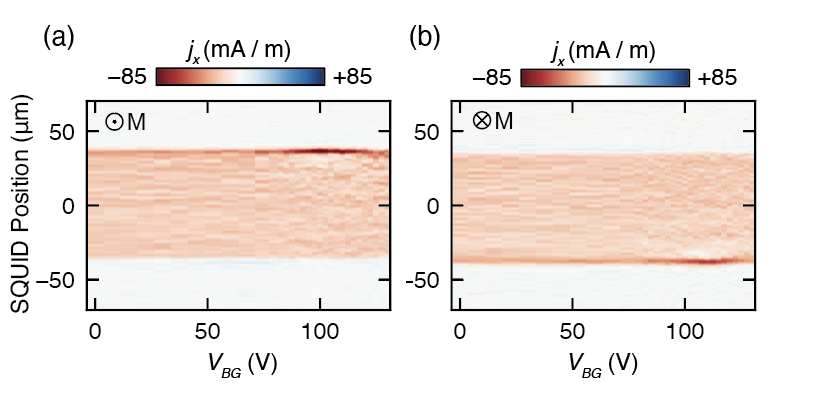}
    \caption{(a) Back-gate dependence of the current distribution acquired by scanning along a line between the two black indicators in Fig. 2a, located between the device voltage probes,  with the sample magnetized out-of the plane. (b) Same as (a) with the sample magnetized into the plane.}
    \label{fig:fig3}
\end{figure}

To understand our observations, we introduce a model, depicted schematically in Fig. 4a, which includes contributions to the current distribution from both sample interior and edge. Within this model, gradients in the electrochemical potential drive currents within the sample bulk, and shifts in the chemical potential within the channel modulate the sample magnetization, 
\begin{equation}
    \mathbf{j}_{tot} =  \mathbf{j}_{\phi} + \nabla \times \mathbf{M} \left( \mu_{ec} \right),
    \label{eq:currents}
\end{equation}
where $\mathbf{M}\left( \mu_{ec} \right) = M_{z}\left( \mu_{ec} \right) \hat{z}$ is the magnetization. Given that both changes in the current distribution and changes in the sample magnetization generate stray magnetic fields, our imaging technique is sensitive to both contributions but cannot directly distinguish between them.

 In the following, we first develop a phenomenological description of the edge currents using the model in Eq. \ref{eq:currents}, and compare it to the data. We defer discussing possible microscopic origins for the dependence of the magnetization on chemical potential to the conclusion of the paper.

 To compare the model to the experimental current distributions shown in Figs. 2 and 3, we combine measurements of the magnetization change in response to a uniform shift in the sample electrochemical potential induced by the top gate with simulations of the electrochemical potential distribution generated by driving a transport current through the channel. To characterize the change in the sample magnetization in response to a uniform shift in the channel chemical potential, we modulated the top gate with a sinusoidal excitation of \SI{240}{\milli\volt} at a frequency of \SI{140}{\hertz} and detected the magnetic response of the sample with the SQUID. In Fig. 4b, we show the reconstructed change in the sample magnetization acquired using this mode of measurement. In previous work \cite{ferguson2023direct}, we used this data as an indicator of local band filling motivated by the change in signal magnitude as we tune through the QAH regime. The changes in magnetization induced by the top gate excitation may equivalently be interpreted as a change in the current distribution in the channel \cite{casola2018review}, given by $\delta \mathbf{j}^M = \nabla \times \left( \delta \mathbf{M} \right)$, where $\delta \mathbf{j}$ and $\delta \mathbf{M}$ are respectively the gate-induced shifts in the current distribution and magnetization. In Fig. 4c, we show $\delta j^M_x /\delta V_{TG}$, the x-component of $\delta \mathbf{j}$ normalized by the amplitude of the top gate excitation. The response of the magnetization in Fig. 4b is uniform across the channel, and therefore the corresponding current density resembles a circulating current along the sample edges. The magnitude of the response is most pronounced in the vicinity of the magnetic exchange gap.

\begin{figure}
    \centering
    \includegraphics[width=0.5\textwidth]{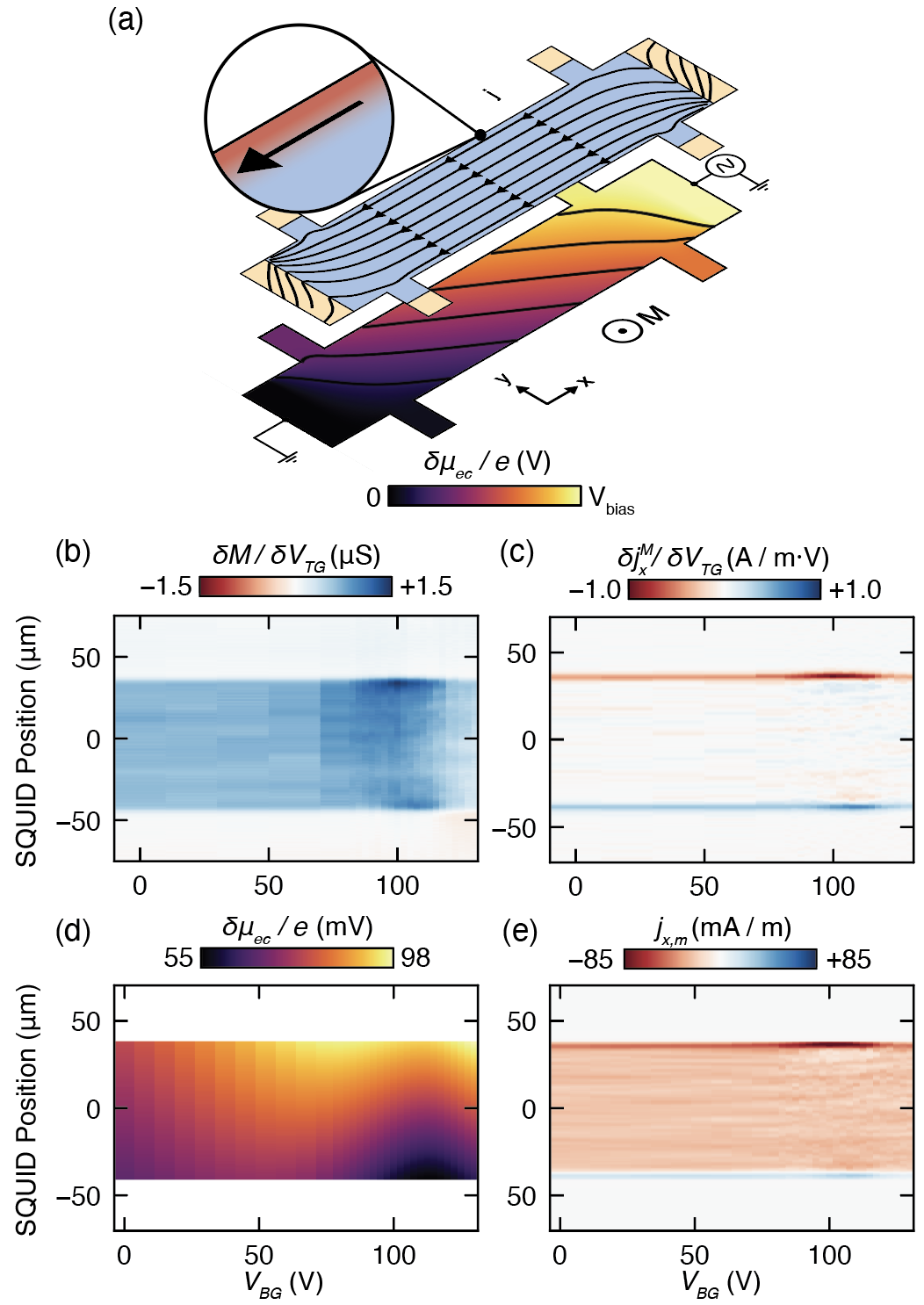}
    \caption{(a) Illustration of the current density (top) and electrochemical potential distribution (bottom) in response to an applied bias current.
    (b,c) Changes induced by a top gate excitation, $\delta V_{TG}$, in (b) the channel magnetization, $\delta M / \delta V_{TG}$, and (c) equivalent current density, $\delta j_x^M / \delta V_{TG}$, as a function of $V_{BG}$. (d) Back gate dependence of the electrochemical potential profile across the width of the channel extracted from $\rho_{xx}$ and $\rho_{xy}$ measurements under the same bias conditions as the magnetic imaging data in Fig. 3. (e) Current density as a function of $V_{BG}$ predicted by our model, calculated using the data in (c) and the potential profiles in (d).}
    \label{fig:fig4}
\end{figure}

 While modulating the top gate generates a uniform shift in the electrochemical potential, an external bias current through the channel induces a non-uniform distribution within the channel. We model the channel as a conductor with a spatially uniform resistivity tensor and assume that the electrostatic potential satisfies the two-dimensional Laplace equation \cite{thouless1985field}. Using the resistivities $\rho_{xx}$ and $\rho_{xy}$ measured under the same bias conditions as the current imaging data, we calculated the electrochemical potential profile along the line the SQUID scans (Fig. 4d). At this location, an average shift in $\mu_{ec}$ results from the longitudinal resistance between the line-cut position and the grounded contact, $R_{xx} = \left( l / w \right) \rho_{xx}$, where $l$ is the distance from the contact, and $w$ is the total channel length. The Hall effect generates a linear $\mu_{ec}$ drop across the channel, $\delta \mu_{ec} = R_{xy} I_b \left(y/w \right)$ with $y$ the position along the channel width measured from the bottom edge.

Using the experimental $\delta j_x^M / \delta V_{TG}$ in Fig. 4b and the $\mu_{ec}$ profile determined from the resistivity in Fig. 4d, we calculate a model current distribution $j_{x,m}$ using the $\hat{x}$ component of the current density in Eq. \ref{eq:currents},
\begin{equation}
    j_{x, m} = j_{uni} + \left( \frac{\delta j_x^M}{\delta V}_{\!\!TG} \right) \cdot \delta \mu_{ec}
\end{equation}
where $j_{uni}$ is uniform across the channel and accounts for contributions from electrostatic potential gradients. The result is shown in Fig. 4e, which we compare to Fig. 3a. The model captures the main features of the experimental current distributions, including the enhanced current at the top edge of the sample, counter-flowing current at the bottom edge, and the $V_{BG}$ dependence of the current amplitudes. Importantly, the calculation in Fig. 4d is not a fit but a direct comparison between the experimentally determined $\delta M / \delta V_{TG}$, resistivity tensor, and current distribution. Additional comparisons under different magnetization and biasing conditions are presented in Supplemental Information Section \ref{si_additional_data}. We emphasize that both the changes in sample magnetization in response to the top gate excitation as well as the changes in magnetization in response to a bias current in the channel are small compared to the total magnetization of the sample. By comparing the amplitude of the flux coupled into the SQUID by the static magnetization, $M_o$, of the sample to the amplitude flux signal generated by modulating $V_{TG}$, we estimate $\delta M / M_o < 3\times10^{-3}$ for the measurement conditions investigated in this work.

A natural question is the origin of the observed dependence of $M$ and corresponding $j^M$ on the chemical potential. Quantum Hall systems are predicted to exhibit a chemical potential dependent magnetization, or equivalently, currents at the sample edges which depend on the chemical potential in the channel \cite{widom1982thermodynamic, buttiker1988absence, streda1983thermodynamic, zhu2020voltage}. The relationship between the sample magnetization and the chemical potential, 
 \begin{equation}
 \frac{\delta I}{\delta \mu} = \frac{\sigma_{xy}}{e}, 
 \label{eq:current}
 \end{equation}
 is predicted to be quantized when the chemical potential is tuned into the gap. Although our model is motivated by the chemical potential dependent currents associated with topologically protected edge states in MTIs, changes in the chemical potential in \CrBST{} may also modify the static magnetization of the Cr dopants \cite{ferguson2023direct}. 

To estimate the contribution from the chiral edge states to the current distribution, we assume that their chemical potential equilibrates locally with the bulk electrochemical potential. The shift in the chemical potential of the edge states, $\delta \mu_{ch}$, is then related through the quantum capacitance, $C_{Q}$, and geometric capacitance of the sample, $C_{G}$ to a shift in the electrochemical potential $\delta \mu_{ec}$ of the bulk:
\begin{equation}
    \delta \mu_{ch} = \frac{C_{Q}}{C_{T}} \delta \mu_{ec},
    \label{eq:mu_relation}
\end{equation}
where $C_{T}$ is the total capacitance given by $1/C_{T} = 1/C_{G} + 1/C_{Q}$.  
From this, we can obtain the change in current flowing in the edge state in response to the top gate modulation as,
\begin{equation*}
    \delta I 
    = \frac{C_Q}{C_T} \sigma_{xy} \cdot \delta V_{TG}, 
\end{equation*}
where we combined Eq. \ref{eq:current} and Eq. \ref{eq:mu_relation}, and use $\delta \mu_{ec}/e = \delta V_{TG}$.

Estimating $\delta I$ requires knowledge of the quantum capacitance $C_Q$. To estimate $C_Q / C_G$, we use the width of quantum anomalous Hall plateau measured in the low-temperature, low-bias limit \cite{rosen2022measured}. Since tuning through the magnetic exchange gap typically requires $\sim \SI{30}{\volt}$ applied to the backgate, we estimate that $C_Q / C_G < 0.01$ in our sample (See Appendix \ref{estimating_c} for details) . This leads to an approximate upper  bound of 90 nA for $\delta I$ given the applied top gate modulation of 240 mV. The observed $\delta M/\delta V_{TG}$ and corresponding $\delta j^M/\delta V_{TG}$ exceed this estimated contribution from chiral edge currents by a factor of $\sim 4-5$ (see Supplementary Information Section \ref{si_dMdV}), which suggests that gate-induced modulation of the magnetization of the Cr-dopants contributes substantially to the observed signal. This also suggests that that the edge currents observed in images of the current distribution in response to an applied bias current are predominantly originating from changes in the Cr-dopant magnetization. These observations are consistent with electrical transport experiments on MTIs \cite{rosen2022measured}, in which the contribution of conduction through the chiral edge states was estimated to be small.


Our estimates for the strength of the edge current signal are consistent with our previous measurements in the low-bias limit, where our device exhibits the QAHE \cite{ferguson2023direct}. In both the quantized and breakdown regimes, the shift in chemical potential of the edge state is determined by $C_Q$ and $C_G$. As long as the chemical potential of the edge states equilibrates with the bulk, the ratio $C_Q/C_T$ determines the fraction of the total source-drain current which is expected to be carried by the edge states. We can establish an upper bound for the edge state contribution the current distribution by assuming that all of the edge current observed in the $\delta M / \delta V_{TG}$ data is due to the chiral edge states. Even in this extreme limit, the small $\sim \SI{15}{\nano\ampere}$ breakdown current of our sample leads us to predict that it supports edge currents of $\sim \SI{1}{\nano\ampere}$ or smaller when the transport coefficients are quantized. A detailed description of the relationship between the edge current amplitudes and the experimentally determined transport coefficients is provided in the Appendix: Modeling the Current Distribution. Although currents this small are below the sensitivity of our measurement \cite{ferguson2023direct}, experiments on devices with larger breakdown currents, \cite{fox2018part} or scanning probe magnetometry with greater current sensitivity, will enable the investigation of edge currents in the quantized transport regime.


Finally, we note that the abrupt change in the ratio between the geometric and quantum capacitances per unit area at the end of the top gate provides a qualitative explanation for the strip of current running vertically along the $\hat{y}$ direction along the boundary of the top gate in Fig. 1c. Underneath the top gate, the geometric capacitance per unit area is substantially larger than in the un-gated regions of the sample, making the electrochemical potential distribution generated by the bias current more effective at modulating the chemical potential dependent magnetization $\mathbf{M}(\mu)$. At the boundary of the top gate, the corresponding  change in the magnetization can be interpreted as a shift in the current density in the $\hat{y}$ direction, $j_y^M$. In the narrow region between the end of the top gate and the current contact, we expect the electrochemical potential distribution to generate smaller shifts in the chemical potential and correspondingly weaker changes in $\mathbf{j}^M$.

Throughout this work we assumed that the edge state chemical potential equilibrates locally with the sample bulk. This assumption may break down at lower bias currents where quantization of the conductivity tensor is recovered, and at smaller length scales than we investigate here, where electrons may propagate ballistically along the sample edges. Current imaging in cleaner samples with a higher spatial resolution is therefore a promising route to observing ballistic edge transport in MTIs.

Our experiments and modeling highlight some of the challenges in disentangling electronic transport signatures in MTIs. Although the current distributions we observe have some qualitative similarities to those predicted for topologically protected edge states, additional information, particularly about the quantum and geometric capacitance of the sample, is required to quantitatively isolate the contribution of the edge states to transport. This is a limitation shared by any attempt to measure edge state transport in MTI systems with magnetic imaging. 

We measured the current distribution in a quantum anomalous Hall insulator driven into breakdown by a large bias current. We identified edge currents which we interpreted through a model of the current distribution in topological insulators which includes contributions from both electrostatic potential gradients as well as the chemical potential dependent magnetization. Future work on higher quality samples will test the validity of this model in the quantized transport regime. Most importantly, our work highlights both the complexity of electronic transport in MTIs as well as the utility of local probes in disentangling different contributions to their electronic transport signatures.

\begin{acknowledgments}
Work at Cornell University was primarily supported by the U.S. Department of Energy, Office of Basic Energy Sciences, Division of Materials Sciences and Engineering, under award DE-SC0015947 with additional support from a New Frontier Grant awarded by Cornell University's College of Arts \& Sciences. Sample synthesis and fabrication at Penn State was supported by the Penn State 2DCC-MIP under NSF Grant nos. DMR-1539916 and DMR-2039351.
\end{acknowledgments}

\appendix
\section{Sample Growth and Fabrication}
We used a VEECO 620 molecular beam epitaxy (MBE) system to grow heterostructures comprised of 3 quintuple layer (QL) \CrBST{} - 5QL \BST{} - 3QL \CrBST{} ~on \ce{SrTiO3}  (111) substrates (MTI Corporation). The Cr composition is nominal (based on past calibrations). The \ce{SrTiO3} substrates were cleaned using deionized water at \SI{90}{\celsius} for 1.5 hours and thermally annealed at \SI{985}{\celsius} for 3 hours in a tube furnace with flowing oxygen gas. The substrate was out-gassed under vacuum at \SI{630}{\celsius} for 1 hour and then cooled down to \SI{340}{\celsius} for the heterostructure growth. When the temperature of substrate was stable at \SI{340}{\celsius}, high-purity Cr (5N), Bi (5N), Sb (6N), and Te (6N) were evaporated from Knudsen effusion cells to form the heterostructure. The desired beam equivalent pressure (BEP) fluxes of each element and the growth rate were precisely controlled by the cell temperatures. The BEP flux ratio of Te/(Bi + Sb) was kept higher than 10 to prevent Te deficiency. The BEP flux ratio of Sb/Bi was kept around 2 to tune the chemical potential of the heterostructure close to the charge neutrality point. The heterostructure growth rate was ~0.25 QL/min, and the pressure of the MBE chamber was maintained at $2 \times 10^{-10}$ mbar during the growth.

After the growth, the heterostructures were fabricated into a \SI{200}{\micro \meter} $\times$ \SI{75}{\micro \meter} Hall bar and a two-terminal sample using photolithography. The shape of the samples was defined by Argon plasma etching. After etching, 10 nm Cr/60 nm Au were deposited outside the active area of the Hall bar to make electrical contact. The top gate was fabricated by depositing a 40 nm \ce{Al2O3} layer by atomic layer deposition across the entire sample and evaporating a 10 nm Ti/60 nm Au layer patterned by optical lithography.

\section{Scanning SQUID Microscopy}
Magnetic imaging was performed with a sample cooled below \SI{50}{\milli\kelvin} in a cryogen free dilution refrigerator, described elsewhere \cite{low2021scanning}. The scanning SQUID sensor has the same gradiometric layout as described in Ref. \cite{huber2008gradiometric} with a \SI{1.5}{\micro\meter} inner-diameter pickup loop. The SQUID is coupled to a SQUID-array amplifier mounted on the mixing chamber plate of the dilution refrigerator. We use a home-built piezoelectric scanner to scan the SQUID $\sim$\SI{1}{\micro\meter} above the sample surface. To measure the flux signal produced by the current distribution in the sample, we excite one of the sample contacts with a sinusoidal current at a frequency of \SI{140.5}{\hertz}. The current contact at the opposite end of the Hall bar is grounded, and the voltage probes are used to simultaneously recorded the longitudinal and Hall voltages. The flux coupled into the SQUID pickup loop is then detected using lock-in techniques. To image the \dMdV{}, The source and drain contacts of the Hall bar are grounded and a lock-in amplifier is used to apply a sinusoidal excitation at a frequency of \SI{140.5}{\hertz} to the sample top gate with an RMS amplitude of \SI{240}{\milli\volt}. The SQUID signal is then demodulated with a lock-in amplifier. 

\section{Current Reconstruction} \label{current_reconstruction}
Given that sample thickness is more than an order of magnitude smaller than both the SQUID pickup loop radius and scan height, we treat the current density as two-dimensional for the purposes of our analysis. The magnetic flux $\Phi(x, y)$ at lateral position $x,y$ and height $z$ above the sample detected by the SQUID is then given by the convolution of the SQUID point spread function, $K_{\textrm{PSF}}$, and the appropriate Biot-Savart kernel, $K_{\textrm{BS}}$,
\begin{equation}
    \Phi(x, y) = K_{\textrm{PSF}}(x, y) \ast K_{\textrm{BS}}(x, y) \ast g(x, y).
    \label{eq:forward_problem}
\end{equation}
Here $\ast$ denotes a convolution,
\begin{equation}
    f(x, y) \ast h(x, y) =
    \int dx' dy' f(x', y') h(x' - x, y' -y).
\end{equation}
The current stream function, $g(x,y)$, determines the two-dimensional current density through,
\begin{equation}
    \vec{j}(x, y) = \nabla \times [g(x, y) \hat{z}].
\end{equation}
In two dimensions, $K_{\textrm{BS}}$ is given by,
\begin{equation}
    K_{\textrm{BS}} = 
    \frac{\mu_o}{2\pi} 
    \frac{2z^2 - x^2 - y^2}{(x^2 + y^2 + z^2)^{5/2}}.
\end{equation}

We extract $K_{\textrm{PSF}}$ from images of superconducting vortices acquired using a nominally identical SQUID. Reconstruction of $g(x, y)$ from a measured image $\Phi(x, y)$ including experimental noise is a deconvolution problem, which requires regularization to avoid the amplification of high spatial frequency noise. Here, we write the problem as a linear system of equations which can then be solved directly. We combine $K_{\textrm{PSF}}$ and $K_{BS}$ into a single linear operator $M$ such that eq. \ref{eq:forward_problem} can be written as $\Phi = Mg$, where now $g$ is a vector with length $n$ equal to the number of pixels in an image and $M$ is a $n \times n$ matrix. Given a suitably chosen regularization operator $\Gamma$ which penalizes solutions that include high-frequency ringing, and a regularization strength $\sigma$, we search for the $g^\ast$ that satisfies,
\begin{equation}
  g^\ast = \mathrm{min}_g \left[ \frac{1}{2}||Mg - \phi||^2 + \sigma^2 ||\Gamma g||^2 \right].
\end{equation}
$g^\ast$  can be found by solving the linear equation,
\begin{equation}
    (M^T M + 2 \sigma^2\Gamma^T \Gamma) g = M^T \phi.
\end{equation}
$M^T$ and $\Gamma^T$ are the pseudo-inverse of $M$ and $\Gamma$ respectively. In practice, we do not directly calculate the elements of $M$, but instead calculate the convolution $Mg$ using Fast Fourier Transforms. We approximate $M^T$ using the Wiener filter, and choose the discrete Laplace operator as our regularization operator $\Gamma$. 

For the one-dimensional line cut data, we utilize the same methods described above in one dimension. In this case, the SQUID point spread function and Biot-Savart kernel are integrated along one axis to form an effective 1D point spread function. 

\section{Electronic Transport Characterization}
Throughout this work, global electronic transport data and local magnetic imaging data were acquired using the same experimental configuration. Electrical connections between room temperature and the sample were provided by attenuating thermocoax cables. We use the reference output of a Stanford Research Systems SR830 lock-in amplifier to excite one current contact of our Hall bar with a sinusoidal excitation at a frequency of \SI{140.5}{\hertz} while leaving the other current contact grounded. The longitudinal and Hall voltages are amplified by Stanford Research Systems SR540 preamplifiers before being demodulated by two lock-in amplifiers. The carrier density in the sample was tuned using the sample back gate formed by the \STO{} substrate. During electrical transport characterization and current density imaging, the top gate was grounded.

\section{Estimating $C_G/C_Q$}
\label{estimating_c}
An estimate of the role that chemical potential dependent edge currents play the current distribution in our sample requires knowledge of the quantum capacitance $C_Q$. We follow the approach outlined in Rec. \cite{rosen2022measured} and compare the width of the QAH plateau to as a function of the back gate voltage to perform a rough estimate of $C_G / C_Q$. MTI samples similar to ours have magnetic exchange gaps of \SI{30}{\milli\electronvolt}. Tuning our sample across the QAH plateau in our sample requires $\sim \SI{30}{\volt}$ on the \ce{SrTiO3} back gate. In our sample, $C_G$ is dominated by the top gate, with $C_G \approx C_{TG} \approx 6 \cdot C_{BG}$ extracted from monitoring the resistivity while performing sweeps of both $V_{TG}$ and $V_{BG}$. The width of the QAH plateau in our sample therefore spans $\sim \SI{5}{\volt}$ on the top gate. We estimate $C_G/C_Q \approx \SI{30}{\milli\volt} / \SI{5}{\volt} = 0.006$.

\section{Modeling the Current Distribution}
Here we derive analytical expressions for the edge current amplitude in a magnetic topological insulator in terms of the sample capacitance and resistivity tensor. To model the current distribution, we consider a top-gated Hall bar with the with a uniform conductivity tensor.

We first consider applying a source-drain bias between the two large contacts on the left and right sides of the device. In steady-state, the electrostatic potential distribution, $\phi$, will satisfy the two-dimensional Laplace equation so that charge does not accumulate in the channel \cite{thouless1985field, rosen2022measured}. The re-distribution of charge which generates $\phi$ will also change the occupation of the edge states, generating an edge current in response.

The chemical potential shift, $\delta \mu_{ch}$ at the sample edge state due to an electrostatic potential shift $\delta \phi$ is given by \cite{buttiker1988absence, halperin1982quantized},
\begin{equation}
    \frac{\delta \mu_{ch}}{e} = \left( \frac{C_G}{C_Q} \right) \delta \phi,
    \label{eq:mu_V}
\end{equation}
where $C_G$ is the geometric capacitance, $C_Q$ is the quantum capacitance and $e$ is the electron charge. This chemical potential shift is in turn related to a change in the edge state current,
\begin{equation*}
    \delta I_{edge} = \left( \frac{e}{h} \right) \delta \mu_{ch} = \left( \frac{e^2}{h} \right) \left( \frac{C_G}{C_Q} \right) \delta \phi,
\end{equation*}
where we have assumed for simplicity that the Chern number $C = 1$ for our system.

To relate the measured potential drops to the electrostatic and chemical potential shifts in the device, we divide the total electrochemical potential shift $\delta \mu_{ec}$ into an electrostatic and chemical potential part, 
\begin{equation}
    \delta \mu_{ec} = \delta \mu_{ch} + e \delta \phi.
    \label{eq:mu_ech}
\end{equation}
Equations \ref{eq:mu_V} and \ref{eq:mu_ech} may be combined to yield expressions for $\delta V$ and $\delta \mu_{ch}$ in terms of $\delta \mu_{ec}$ and the sample capacitances,
\begin{align}
    \delta \phi & = \left( \frac{C_T}{C_G} \right) \frac{\delta \mu_{ec}}{e}\\
    \delta \mu_{ch} & = \left( \frac{C_T}{C_Q} \right) \delta \mu_{ec},
    \label{eq:mu_ch_mu_ech}
\end{align}
where $C_T$ is the total capacitance of the sample given by $1/C_T = 1/C_G + 1/C_Q$.

The voltmeters connected to our device report $V_{ij}$, which provides a measurement of the difference in electrochemical potential between a pair of contacts,
\begin{equation*}
    V_{ij} = \frac{\mu_{ech, i} - \mu_{ech, j}}{e}.
\end{equation*}

We determine the resistivity tensor of the device by measuring $V_{ij}$ in a longitudinal ($V_{xx}$) and Hall ($V_{xy}$) geometry,
\begin{align*}
    \rho_{xx} &= \frac{V_{xx}}{I_{bias}} \frac{w}{l} \\
    \rho_{xy} &= \frac{V_{xy}}{I_{bias}}.
\end{align*}

We determine the downstream edge current $I_{d}$ and back-scattered, upstream edge current $I_{u}$ by calculating the chemical potential shift at the downstream ($\mu_{ch, d}$) and upstream ($\mu_{ch, u}$) edges of the sample. $I_{u}$ and $I_{d}$ can then be related to the electrochemical potential shifts at the sample edges using \ref{eq:mu_ch_mu_ech},
\begin{align*}
    I_{d} &= \left( \frac{e}{h} \right) \mu_{ch, d}
    = \left( \frac{e}{h} \right) \left( \frac{C_T}{C_Q} \right) \mu_{ech, d},
    \\
    I_{u} &= \left( \frac{e}{h} \right) \mu_{ch, u}
    = \left( \frac{e}{h} \right) \left( \frac{C_T}{C_Q} \right) \mu_{ech, u}.
\end{align*}
The shift in the electrochemical potential in response to a bias current is in turn related to the electronic transport coefficients,
\begin{align}
    I_d &= \left( \frac{e^2}{h} \right) \left( \frac{C_T}{C_Q} \right) \left( \rho_{xx} \frac{l_d}{w} + \left|\rho_{xy}\right|\right) I_{bias} \label{eq:app1}
    \\ 
    I_u &= \left( \frac{e^2}{h} \right) \left( \frac{C_T}{C_Q} \right) \left( \rho_{xx} \frac{l_d}{w} \right) I_{bias},
    \label{eq:app2}
\end{align}
Where $l_d$ is the distance from where the edge currents are measured to the drain contact. The net current transported through the sample along the edge $I_{edge}$, is given by the difference between the downstream and upstream edge currents, 
\begin{equation}
    I_{edge} = I_{d} - I_{u} = 
    \left( \frac{e^2}{h} \right)
    \left( \frac{C_T}{C_Q} \right)
    \left|{V_{xy}} \right|,
    \label{eq:app3}
\end{equation}
and the ratio between $I_{edge}$ and the total source drain current is can be obtained by dividing out the bias current $I_{bias}$,
\begin{equation}
    \frac{I_{edge}}{I_{bias}} = 
    \left( \frac{C_T}{C_Q} \right)
    \frac{\left| \rho_{xy} \right|}{R_K},
    \label{eq:app4}
\end{equation}
with $R_K$ the von Klitzing constant. The equations \ref{eq:app1} -- \ref{eq:app4} provide a relationship between the local edge current amplitudes and the global electronic properties of the sample which may be determined independently and checked for internal consistency. 


In this analysis we have neglected the edge state density of states, assuming that the density of localized states in the magnetic exchange gap is much larger than the one-dimensional density of states at the sample edges. In principle, we expect the situation to qualitatively change if $\sigma_{xx} = 0$, as predicted for an ideal quantum Hall insulator \cite{thouless1985field}. In this limit, charge cannot diffuse into the sample bulk, and the electrostatic potential distribution no longer satisfies the Laplace equation. Although this situation has been studied extensively theoretically in the context of the integer quantum Hall regime, real magnetic topological insulator samples all exhibit a non-zero $\sigma_{xx}$.


\bibliography{references}

\clearpage
\newpage

\section*{Supplemental Information}

\renewcommand{\thefigure}{S\arabic{figure}}
\renewcommand{\thetable}{S\arabic{table}}
\setcounter{figure}{0}   

\section{Images and simulations of the contact area} \label{si_electrostatics}
\subsection{Model for current injection} 
To model the magnetic flux coupled into the SQUID, we follow the approach of Sample \textit{et al.} \cite{sample1987reverse}. In short, we model our device as grid of nodes where the electrostatic potential $V$ is defined. The nodes in the grid are linked by a network of resistors which support currents driven by the potential differences between neighboring nodes. Within our model, each pair of neighboring nodes are linked by two resistors. The first resistor encodes the effects of the longitudinal conductivity and the current is determined by the longitudinal potential drop between the nodes. In the second resistor, the current between the nodes is proportional to the transverse potential drop between the nodes, encoding the Hall effect. We use the connectivity of the nodes and the values of the resistances to populate a sparse matrix which we then solve for the electrostatic potential distribution $V$ inside the device. This approach produces equivalent results to those obtained with direct numerical solutions of the Laplace equation \cite{rosen2022measured}, or the conformal mapping approach \cite{kirtley1986voltage}.

Once the electrostatic potential distribution is known, we use the conductivity tensor to determine the current distribution using Ohm's law,
\begin{align*}
    j_x &= \sigma_{xx}E_x + \sigma_{xy}E_y \\
    j_y &= \sigma_{xx}E_y + \sigma_{yx}E_x,
\end{align*}
where we have used $\sigma_{xx} = \sigma_{yy}$. To compare the calculated current distributions to our magnetic imaging data, we convolve the current distribution with the point spread function of our imaging. This procedure generates a model $\Phi_{AC}$ image which may then be directly compared to experimental images of $\Phi_{AC}$. 

Within our model, the contacts are defined to have $\sigma_{xy} = 0$ and $\sigma_{xx} \gg h/e^2$. The semiconducting channel is modeled as a material with a uniform conductivity tensor with $\sigma_{xy} \neq 0$. For a particular calculation, we use the measured longitudinal and Hall resistivity to set the Hall angle $\theta_H = \rho_{xy} / \rho_{xx}$ inside the channel. 

An example of these calculations are shown in Fig. \ref{fig:si_hot_spot_motion}(a-d). In Fig. \ref{fig:si_hot_spot_motion}(a, b) we plot the $\Phi_{DC}$ and $\Phi_{AC}$ images acquired at the interface between the Au contact and \CrBST{} channel, reproduced from Fig. 1. In Fig. \ref{fig:si_hot_spot_motion}(c), we show the electrostatic potential distribution, $V$, normalized by the potential $V_b$ applied to the contact, calculated using our resistor network approach. In Fig. \ref{fig:si_hot_spot_motion}(d) we show a model $\Phi_{AC}$ image obtained by convolving the current distribution calculated from Fig. \ref{fig:si_hot_spot_motion}(c) with the point spread function of our imaging technique. We find that the resistor network model does a good job of reproducing the main features of the experimental $\Phi_{AC}$ image.

The resistor network model tends to over-estimate the strength of the magnetic flux signature at the top corner of the contact. This quantitative disagreement likely stems from the enormous current density at the corner of the contact. The measured resistivity of our sample, which we use the constrain $\theta_H$ in our model, depends on the current density in the channel. In our sample geometry, this constraint is expected to be accurate far from the Au contacts, where the current is distributed uniformly over the sample width. At the top corner of the contact, where the current density is enhanced, we expect the sample to be driven further into breakdown causing a suppression of $\rho_{xy}$ and an enhancement in $\rho_{xx}$. We expect $\theta_H$ is locally suppressed where the current density is highest, leading to a partial suppression of the current-focusing effect responsible for the $\Phi_{AC}$ feature at the top of the contact.

Away from the Au contact, the resistor network model predicts a completely uniform current distribution over the width of the \CrBST{} channel. Although this is a good first approximation for the data shown \ref{fig:si_hot_spot_motion}(b), the resistor network model does not reproduce the signatures of edge conduction at the bottom edge of the sample. To capture these details of the current distribution, the complexity of the model must be increased by either introducing non-uniformity into the conductivity tensor of the device or by introducing edge states into the description of the system.

\subsection{Magnetization reversal}
\begin{figure*}
    \centering
    \includegraphics[width=1.0\textwidth]{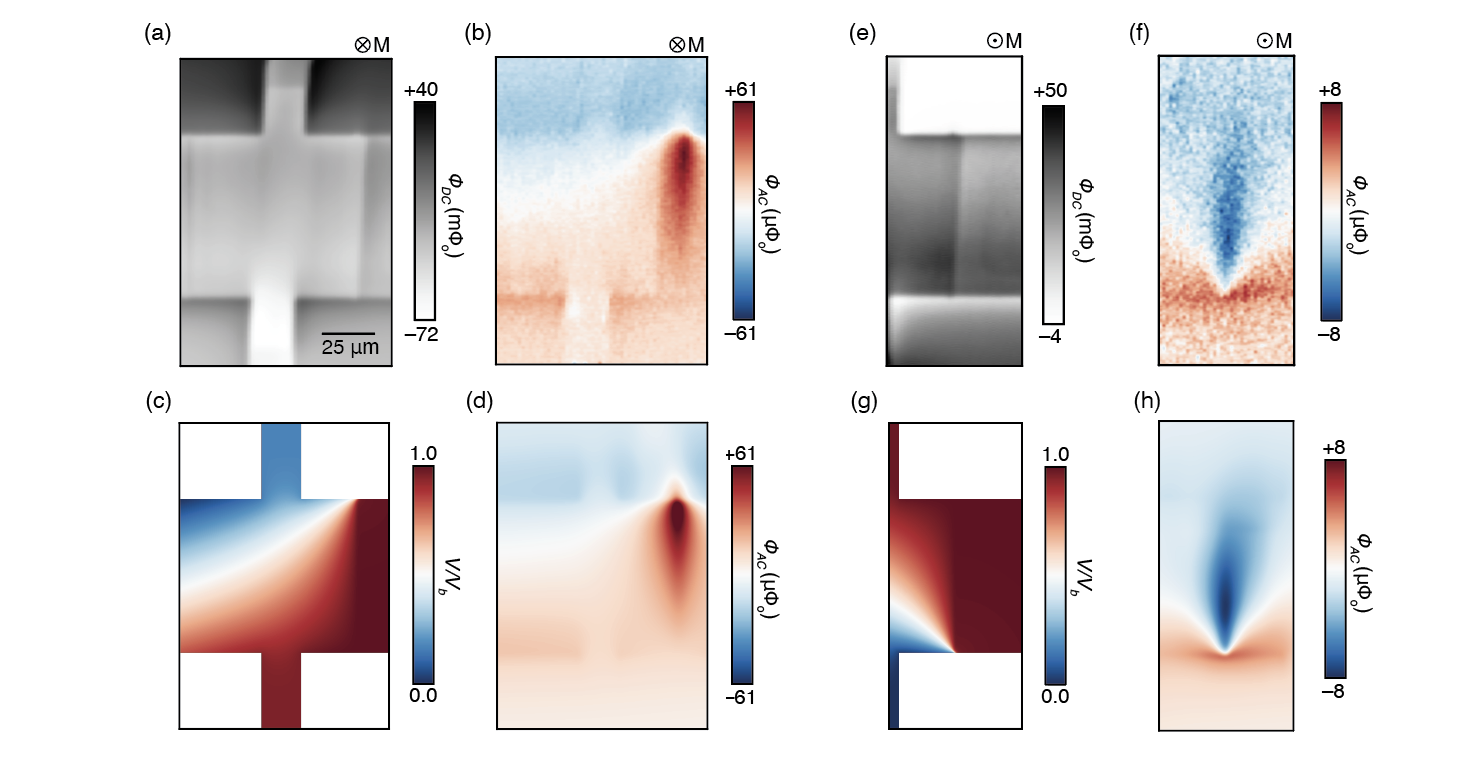}
    \caption{(a) $\Phi_{DC}$ coupled into the SQUID by the magnetization of the sample with the device magnetized into the plane. Reproduced from Fig. 1(b). (b) $\Phi_{AC}$ coupled into the SQUID by the \SI{500}{\nano\ampere} current in the vicinity of the interface between the contact and the \CrBST{} channel. Reproduced from Fig. 1(c). (c) Simulation of the electrostatic potential profile in the vicinity of the contact, normalized by the bias $V_b$ on the contact. (d) Model $\Phi_{AC}$ image generated using the electrostatic potential distribution in (c). (e) Same as (a) with the sample magnetization reversed. (f) $\Phi_{AC}$ coupled into the SQUID by a \SI{60}{\nano\ampere} current in the vicinity of the Au contact. (g) Same as (c) with the sample magnetization reversed. (h) Same as (d) calculated using the potential distribution in (g) and $I_b = \SI{60}{\nano\ampere}$.}
    \label{fig:si_hot_spot_motion}
\end{figure*}

In Fig. \ref{fig:si_hot_spot_motion}(e-h) we show the effects on reversing the sample magnetization on the current injection behavior near the contact. Reversing the magnetization of the sample modified the current distribution in the vicinity of the contact area. In Fig. \ref{fig:si_hot_spot_motion}(e), we show the $\Phi_{DC}$ image acquired after reversing the sample magnetization. The contrast of the image is inverted relative to Fig. \ref{fig:si_hot_spot_motion}(a). The interface between the Au contact and the semiconducting channel is visible as a vertical line in the image. In Fig. \ref{fig:si_hot_spot_motion}(f), we show the $\Phi_{AC}$ image acquired while scanning the field of view in \ref{fig:si_hot_spot_motion}(e), while sourcing a \SI{60}{\nano\ampere} RMS bias current through the contact. In contrast to Fig. \ref{fig:si_hot_spot_motion}(b), where we observed a strong $\Phi_{AC}$ signal at the top corner of the contact, we observe a strong $\Phi_{AC}$ signature at the bottom corner of the contact interface.

To understand this behavior, we repeated the electrostatic potential simulations from Fig. \ref{fig:si_hot_spot_motion}(c), with the sign of the Hall conductivity reversed. In \ref{fig:si_hot_spot_motion}(g), we show the results of this calculation. We find that reversing the sign of the Hall conductivity in the channel causes the strong electric field gradients to appear at the bottom corner of the contact instead of the top corner. To compare this electrostatic potential distribution to our magnetic imaging data, repeated the procedure of Fig. \ref{fig:si_hot_spot_motion}(d), using the gradients in the electrostatic potential and the conductivity tensor to calculate the current distribution and then convolving the current distribution with the point spread function of our imaging technique. We show the results of this calculation in Fig. \ref{fig:si_hot_spot_motion}(h). Our model does a good job of capturing behavior observed in the $\Phi_{AC}$ image in Fig. \ref{fig:si_hot_spot_motion}(f).

\section{Magnetic imaging data used for current reconstruction}

\begin{figure*}
    \centering
    \includegraphics[width=0.5\textwidth]{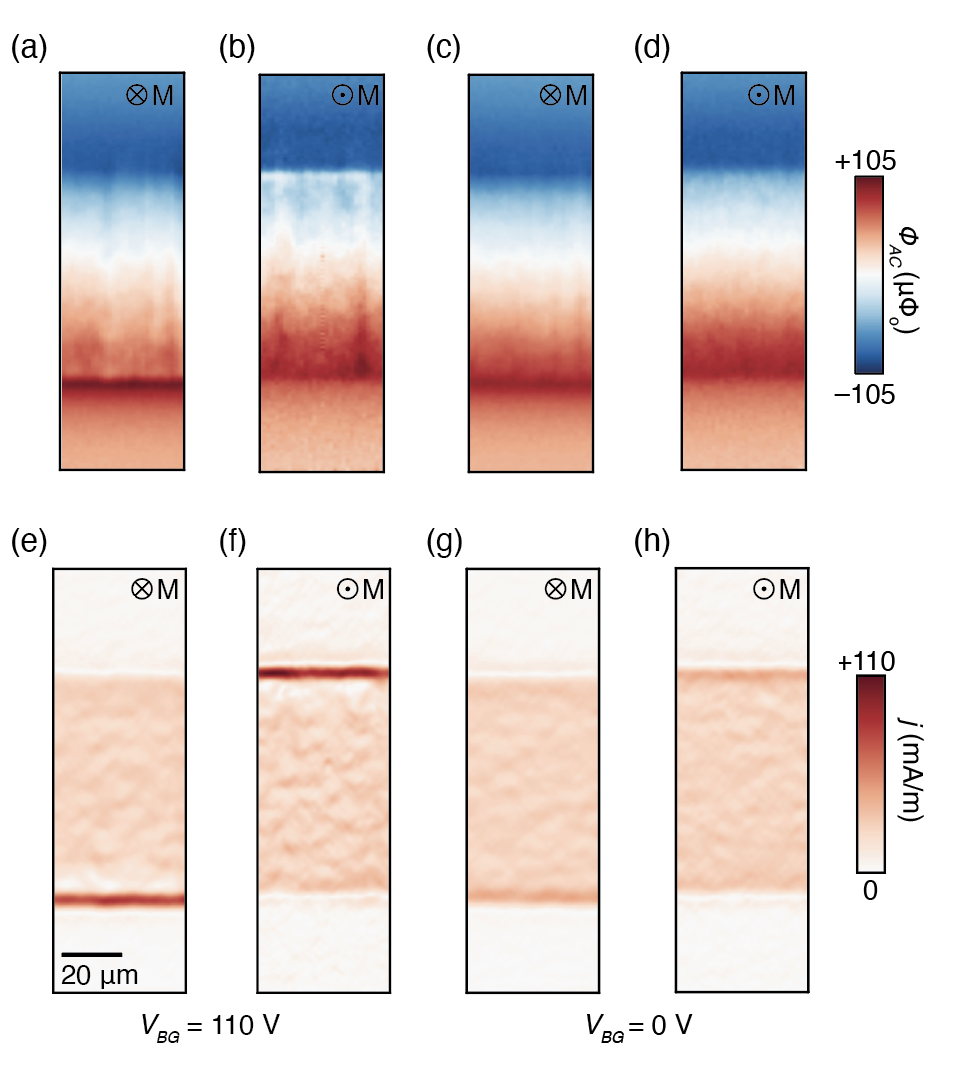}
    \caption{(a) $\Phi_{AC}$ measured between the voltage probes of the \CrBST{} device with the sample magnetized into the plane and gated into the magnetic exchange gap with $V_{BG} = \SI{110}{\volt}$. (b) Same as (a) with the magnetization reversed. (c) Same as (a) with the sample tuned out of the magnetic exchange gap with $V_{BG} = \SI{0}{\volt}$. (d) Same as (c) with the magnetization reversed. (e-f) reconstructed current densities from the magnetic flux images in (a-d), reproduced from Fig 2.
    }
    \label{fig:raw_img_flux}
\end{figure*}

In Fig. \ref{fig:raw_img_flux}, we show the $\Phi_{AC}$ images used the reconstruct the two-dimensional current density defined as $j = \sqrt{j_{x}^2 + j_{y}^2}$ presented in Fig. 2 in the main text. Signatures of edge conduction are observable in the $\Phi_{AC}$ images, where an enhanced current density at the sample edge appears as a sharp gradient in $\Phi_{AC}$ at the edge of the device. In Fig. \ref{fig:raw_img_flux}(a), we observe a strong gradient in the magnetic flux at the bottom edge of the sample, while in Fig. \ref{fig:raw_img_flux}(b), we observe a strong gradient at the top edge of the sample. We find a similar effect in Fig. \ref{fig:raw_img_flux}(c, d), but the flux gradients are the sample edges are smaller than those measured in  Fig. \ref{fig:raw_img_flux}(a, b), indicating that the amplitude of the edge current is smaller outside the magnetic exchange gap with $V_{BG} = \SI{0}{\volt}$, than when the device is tuned into the magnetic exchange gap with $V_{BG} = \SI{110}{\volt}$. 

\begin{figure*}
    \centering
    \includegraphics[width=0.5\textwidth]{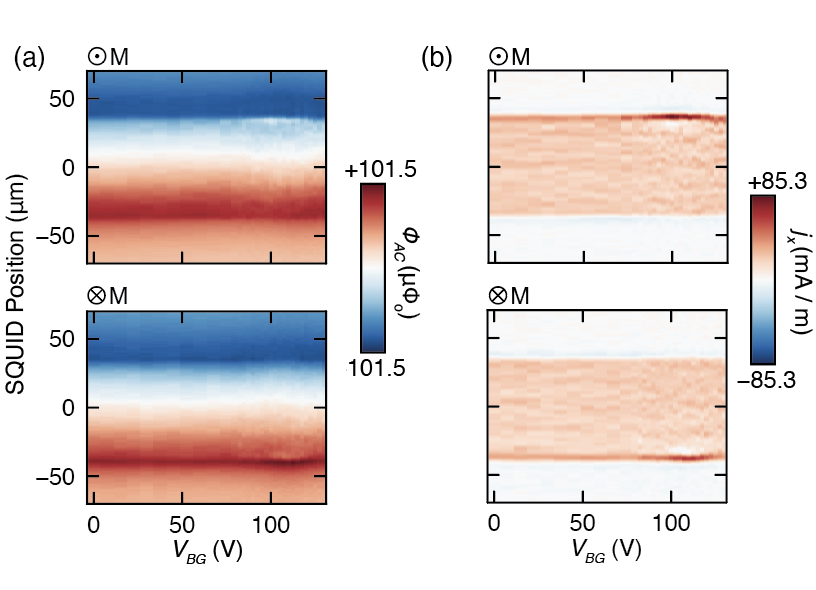}
    \caption{(a) $\Phi_{AC}$ measured over the width of the channel as a function of $V_{BG}$ with the sample magnetized out-of (top) and into (bottom) the plane. (b) Reconstructed current distribution from the $\Phi_{AC}$ in (a), reproduced from Fig. 3.
    }
    \label{fig:raw_lc_flux}
\end{figure*}

In Fig. \ref{fig:raw_lc_flux} we show the $\Phi_{AC}$ linecut data used to reconstruct the current distribution as a function of $V_{BG}$ in Fig. 3. Just as we found for the $\Phi_{AC}$ images (Fig. \ref{fig:raw_img_flux}, strong gradients in $\Phi_{AC}$ at the sample edges in Fig. \ref{fig:raw_lc_flux} (a) are reconstructed as edge currents in  \ref{fig:raw_lc_flux} (b). The strength of the $\Phi_{AC}$ gradient at the sample boundary is strongest when the sample is tuned into the magnetic exchange gap with $V_{BG} = \SI{110}{\volt}$.

\section{Magnetic imaging of a \CrST{} device} \label{si_CRST}

\begin{figure*}
    \centering
    \includegraphics[width=0.5\textwidth]{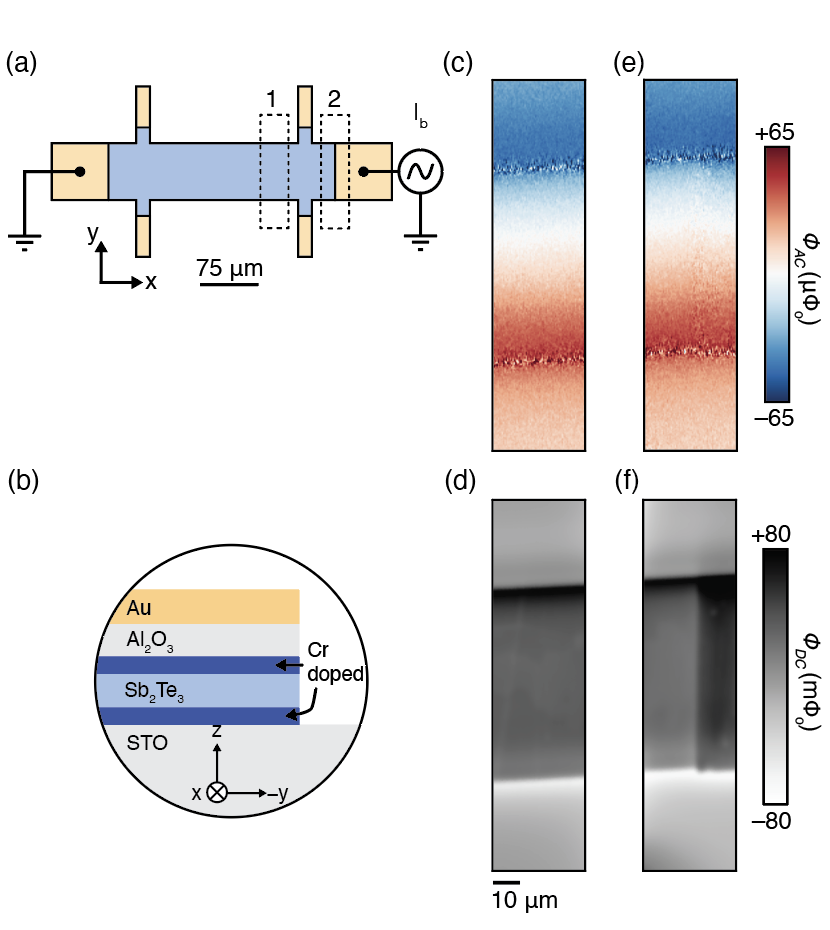}
    \caption{(a) Schematic illustration of the \CrST{} measurement geometry. The dimensions of the sample are identical to the \CrBST{} sample discussed in the main text. (b) Cross-sectional schematic of the \CrST{} sample composition. An undoped layer of $\mathrm{(Bi,Sb)_{2}Te_{3}}$ is sandwiched between two heavily Cr-doped layers of \CrST{}. (c) $\Phi_{AC}$ coupled into the SQUID pickup loop by a \SI{1}{\micro\ampere} RMS bias current in region 1 in from the schematic in (a). The magnetic flux profile is consistent with a current that is uniformly distributed over the sample width. (d) $\Phi_{DC}$ coupled into the SQUID by the sample magnetization in the in region 1. (e) $\Phi_{AC}$ coupled into the SQUID pickup loop by a \SI{1}{\micro\ampere} RMS bias current in region 2 at the interface between the device channel and the gold contact. The magnetic flux profile is consistent with a current that is uniformly distributed over the sample width. The redistribution of the current at the contact interface observed at in \CrBST{} is absent in the \CrST{} device. (f) $\Phi_{DC}$ coupled into the SQUID by the sample magnetization in the in region 2 at the interface between the device channel and the gold contact. 
    }
    \label{fig:CrST_images}
\end{figure*}

We measured a \CrST{} Hall bar with identical dimensions to \CrBST{} described in detail in the main text. The sample composition and geometry are depicted schematically in Fig. \ref{fig:CrST_images}. In \CrST{}, the chemical potential lies deep in the valence band, far below the magnetic exchange gap. The \CrST{} devices exhibit only a small anomalous Hall effect of $\sim \SI{50}{\ohm}$.

In Fig. \ref{fig:CrST_images}(c, d), we show the $\Phi_{AC}$ and $\Phi_{DC}$ images acquired from scanning the SQUID between the voltage probes on the sample while sourcing a \SI{1}{\micro\ampere} RMS bias current through the channel. These images may be directly compared to the raw magnetic flux data acquired on the \CrBST{} sample (Fig. \ref{fig:raw_img_flux}), where we observed an edge current. The sharp gradient in $\Phi_{AC}$ which indicates an excess current density at the sample edge is absent from Fig. \ref{fig:CrST_images}(c). The $\Phi_{AC}$ image in Fig. \ref{fig:CrST_images}(c) is consistent with a current distribution that is uniform over the width of the channel. 

In Fig. \ref{fig:CrST_images}(e, f), we show $\Phi_{AC}$ and $\Phi_{DC}$ images acquired while scanning the SQUID over the interface between the Au contact and \CrST{} channel. The interface between the Au contact and \CrST{} channel is visible in the $\Phi_{DC}$ image. Unlike the \CrBST{} device, we observe no significant features in the $\Phi_{AC}$ image at the interface between the Au contact and \CrST{}. The $\Phi_{AC}$ image in Fig. \ref{fig:CrST_images}(e) is consistent with a uniform current distribution. 

The differences in the current distribution at the contact interface between \CrST{} and \CrBST{} may be understood by considering the large differences in the Hall angle, defined by $\theta_H = \rho_{xy} / \rho_{xx}$, between the two samples. Inside the metal contact, $\theta_H = \SI{0}{\degree}$, and the electric field is parallel to the current density, $\vec{E} \parallel \vec{j}$. When the current is injected into the semiconducting channel, the electric field rotates from being parallel to the current density near the contact interface such that the angle between $\vec{E}$ and $\vec{j}$ is equal to $\theta_H$. 

In \CrBST{}, where $\theta_H \approx \SI{45}{\degree}$ even in breakdown, the local electric field and current distributions are substantially modified at the contact interface. The effect of the change in $\theta_H$ at the interface between the \CrBST{} channel and Au contact is shown in Fig \ref{fig:si_hot_spot_motion}(c, g). In \CrST{}, $\theta_H < \SI{1}{\deg}$, and changes to the electrostatic potential distributions and the local current density at the interface are expected to be small.

\section{Analysis of Additional Current Distributions} \label{si_additional_data}
\begin{figure*}
    \centering
    \includegraphics[width=0.5\textwidth]{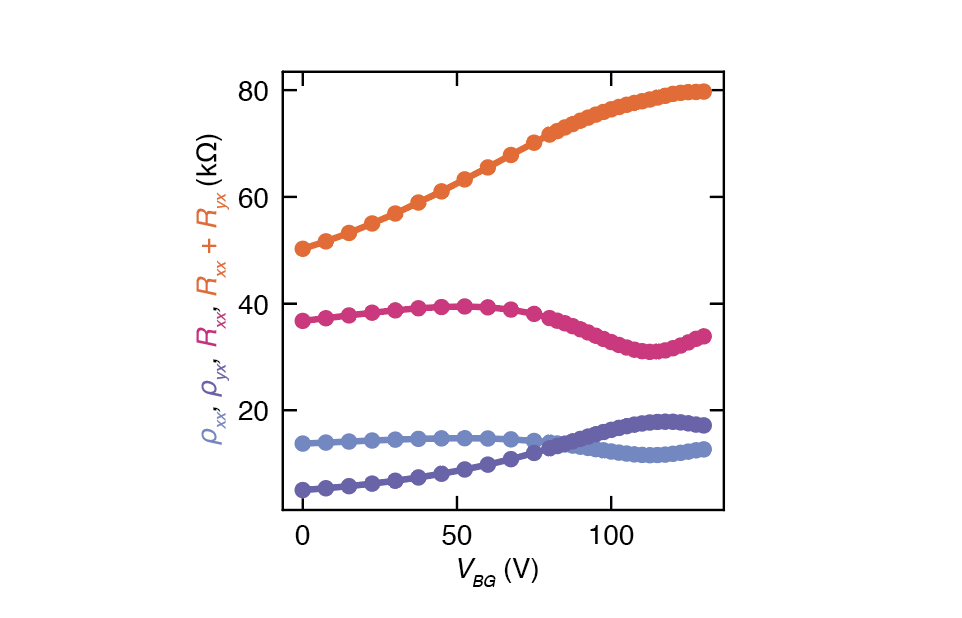}
    \caption{
    $V_{BG}$ dependence of the sample resistivity $\rho_{xx}$ (blue) and $\rho_{yx}$ (purple) with $I_b = \SI{2}{\micro\ampere}$ used to determine the electrochemical potential profiles in Fig. 4c. In pink, $R_{xx} = \left(l / w \right) \rho_{xx}$ where $l = \SI{200}{\micro\meter}$ is the distance from the location where the current distribution is recorded to the grounded contact on the left end of the sample. In Fig 4c, $R_{xx}$ sets the amplitude of $\delta \mu_{ec}$ on the bottom edge of the sample. $R_{xx} + R_{yx}$ (orange) determines the amplitude of $\delta \mu_{ec}$ on the top edge of the sample.
    }
    \label{fig:R_Vbg}
\end{figure*}

In Fig. \ref{fig:R_Vbg} we show the $V_{BG}$ dependence of $\rho_{xx}$ and $\rho_{xy}$ used to characterize the electrochemical potential distribution shown in Fig. 3c. From the measured $\rho_{xx}$, we calculate $R_{xx} = \left(l / w \right) \rho_{xx}$, where $l$ is the distance from the SQUID to the grounded contact on the sample. This $R_{xx}$ determines the shift $\delta \mu_{ec}$ on the bottom edge of the sample. We use the sum $R_{xx} + R_{xy}$ to determine $\delta \mu_{ec}$ on the bottom edge of the sample. 

\subsection{Reversed Magnetization}
\begin{figure*}
    \centering
    \includegraphics[width=0.5\textwidth]{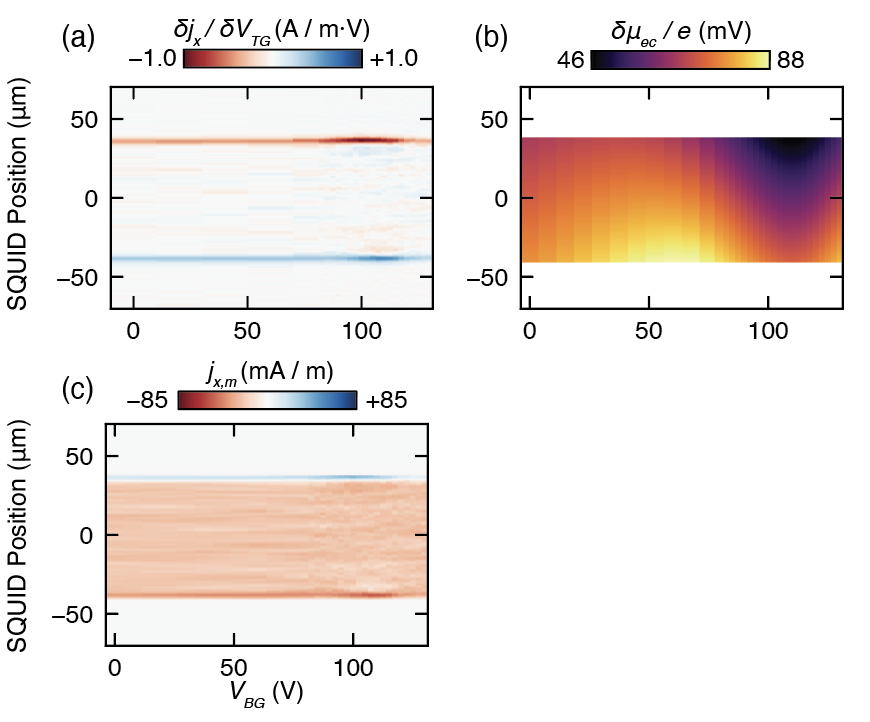}
    \caption{(a) Change in the current distribution in response to a top gate excitation as a function of $V_{BG}$, reproduced from Fig. 4c. (b) Back gate dependence of the electrochemical potential profile across the width of the channel extracted from $\rho_{xx}$ and $\rho_{xy}$ measurements under the same bias conditions as the magnetic imaging data in Fig. 3b. (c) Current distribution as a function of $V_{BG}$ predicted for our model calculated using the data in (a) and the potential profiles in (b).
    }
    \label{fig:model_comp_neg_mag}
\end{figure*}

In Fig. \ref{fig:model_comp_neg_mag} we repeat the analysis of Fig. 4 with the magnetization of the sample reversed, using the data shown in Fig. 3b.

\subsection{Lower Bias Current}
Both $\rho_{xx}$ and $\rho_{xy}$ depend on the bias current $I_b$ driven through the sample. To compare the current distributions in our device to our model for different values of the resistivity, we repeated the measurements and analysis shown in Fig. 4 with a smaller $I_b = \SI{500}{\nano\ampere}$ and show the results in In Fig. \ref{fig:model_comp_neg_mag}.

\begin{figure*}
    \centering
    \includegraphics[width=0.5\textwidth]{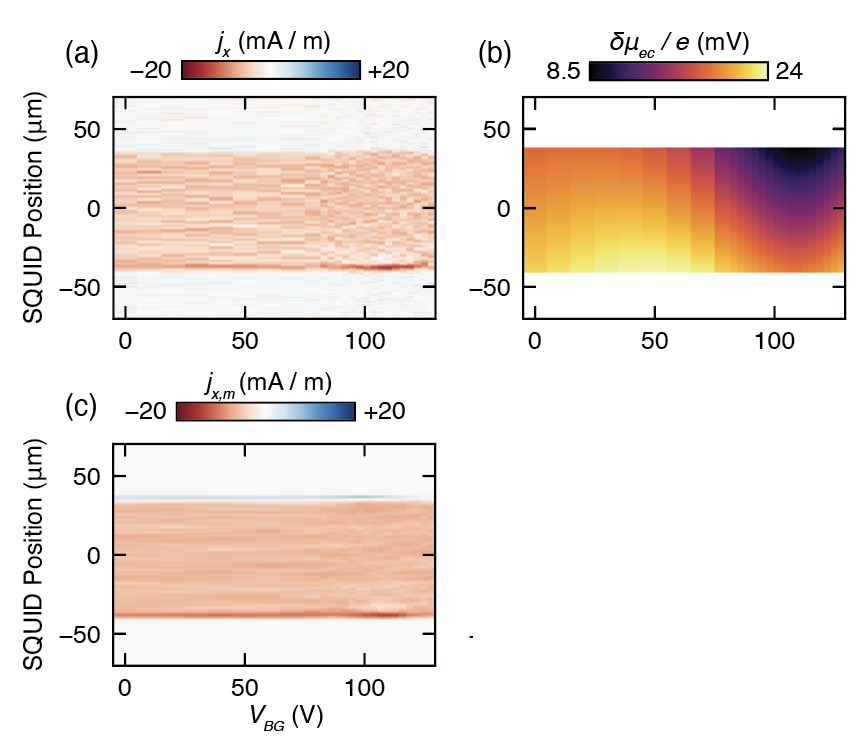}
    \caption{(a) Reconstructed current distribution over the width of the device channel as a function of $V_{BG}$ using a source-drain bias current $I_b = \SI{500}{\nano\ampere}$ (b) Back gate dependence of the electrochemical potential profile across the width of the channel extracted from $\rho_{xx}$ and $\rho_{xy}$ measurements under the same bias conditions as the reconstructed current distributions in (a). (c) Current distribution as a function of $V_{BG}$ predicted for our model calculated using the potential profiles in (b) and the $\delta j_x / \delta V_{TG}$ data in Fig. 4(a). (d) Difference between the model (c) and the experimentally determined current distribution in (a). 
    }
    \label{fig:500_nA_lc}
\end{figure*}

\subsection{Contact Reversal}

\begin{figure*}
    \centering
    \includegraphics[width=0.5\textwidth]{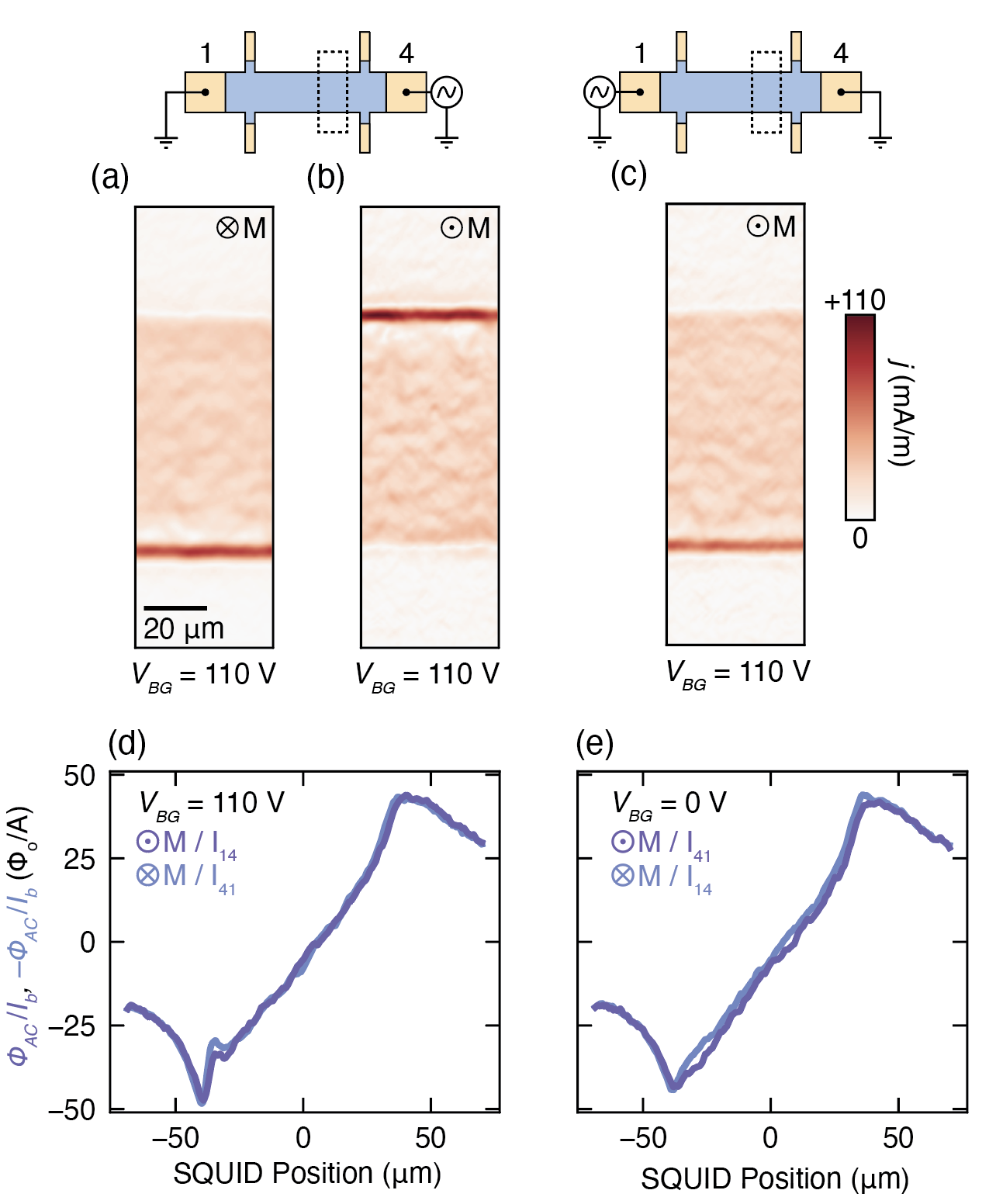}
    \caption{(a) Reconstructed current density with the sample tuned into the magnetic exchange gap and magnetized into the plane. $I_{b} = \SI{2}{\micro\ampere}$. (b) Same as (a) with the magnetization reversed. (c) Same as (b) with the source and drain contacts interchanged. An enhanced current density with a reduced amplitude is observed on the bottom edge of the sample when the SQUID is scanned closer to the grounded contact. (d) Comparison of the SQUID signal with the contact further away from the SQUID grounded (blue) to the SQUID signal with the contact closer to the SQUID grounded (purple). The phase of the of the current relative to the lock-in reference is inverted for the two signals so the blue trace is multiplied by $-1$ to facilitate comparison between the two signals. (e) Same as (d) with the sample gated out of the magnetic exchange gap with $V_{BG} = \SI{0}{\volt}$.
    }
    \label{fig:si_contact_reversal}
\end{figure*}

In our model for the chemical potential dependent edge current, the non-uniform electrochemical potential distribution generated within the channel by a source-drain bias plays a role in determining the contribution to the SQUID signal from changes in the sample magnetization. The longitudinal resistance between the SQUID measurement location and the grounded contact is given by $R_{xx} = \rho_{xx} l / w$, where $l$ is the distance between the SQUID x-position and the grounded contact, and $w$ is the channel width. $R_{xx}$ generates a uniform shift in the electrochemical potential across the sample width when a bias current $I_b$ passes through the channel which depends on the distance between the SQUID and the grounded contact.

In the main text, all measurements were presented with the contact furthest from the SQUID (negative x-direction) grounded and the contact nearest the SQUID excited by the lock-in amplifier. In this configuration, the distance between the SQUID and the grounded contact was $l = \SI{210}{\micro\meter}$ (Fig. \ref{fig:si_contact_reversal}). To test the dependence of the SQUID signal on the distance to the contact, we reversed the source and drain contacts so that the contact nearest to the SQUID was grounded, yielding $l=\SI{75}{\micro\meter}$. In Fig. \ref{fig:si_contact_reversal} (a-c) we show the dependence of the edge current amplitude on the reversal of the contacts. We find that a simultaneous exchange of the source drain contacts combined with a reversal of the sample magnetization produces an enhanced current density on the bottom edge of the sample. When the grounded contact is closer to the SQUID, the amplitude of the edge current is reduced, in qualitative agreement with the expectations of our model.

In Fig. \ref{fig:si_contact_reversal} (d) we show compare SQUID flux signal detected while for both contact configurations. When the contact further away from the SQUID is grounded, we observe a stronger edge current signature at the bottom edge of the sample (blue trace) than when the contact closer to the SQUID is grounded. Inverting the contacts creates a \SI{180}{\degree} phase shift in the current distribution. To directly compare the flux signals, we multiply the blue trace by $-1$.

\section{Estimating the Edge Current Amplitude}
\label{si_dMdV}
\begin{figure*}
    \centering
    \includegraphics[width=1.0\textwidth]{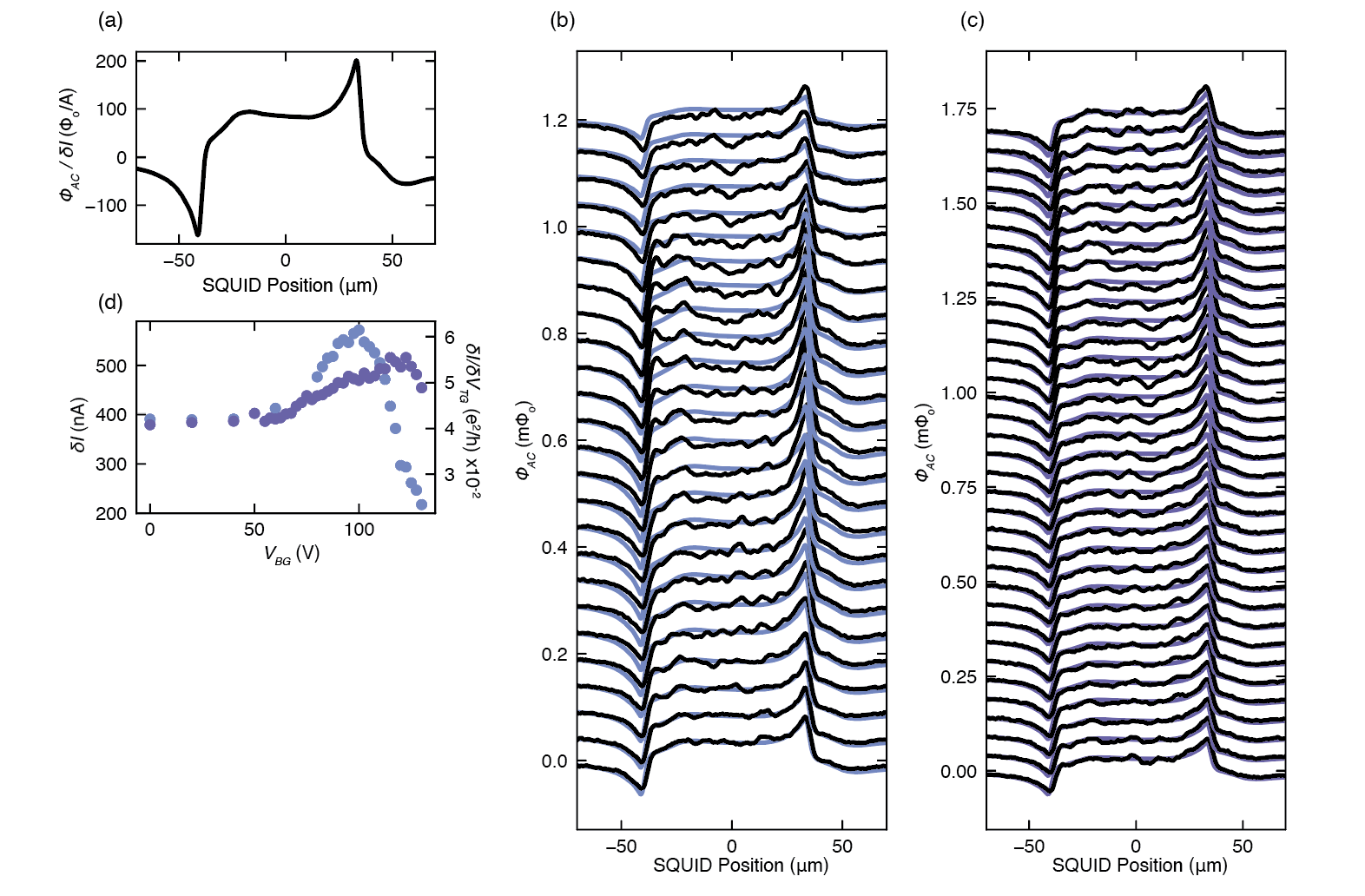}
    \caption{(a) Model magnetic flux profile generated by convolving the SQUID point spread function with a uniform change in the channel magnetization. (b-c) $V_{BG}$ dependence of the flux coupled into the SQUID due to top-gate modulation $\delta \Phi/\delta V_{TG}$ acquired scanning the SQUID along a line over two different positions over the channel. Traces are offset for clarity and arranged in order of increasing $V_{BG}$ with $V_{BG} = 0$ at the bottom. Fits of the model in (a) to the $\delta \Phi/\delta V_{TG}$ data with $\delta M$ as a free parameter are shown in black. (d) $V_{BG}$ dependence of $\delta M/\delta V_{TG}$ extracted from the fits to the flux profiles in (b) (blue) and in (c) (purple).
    }
    \label{fig:raw_dMdV}
\end{figure*}

To make a quantitative comparison between the amplitude of the gate-induced changes in the sample magnetization, or equivalently the currents at the edge of the sample, we fit a line profile of a uniform shift in the sample magnetization to the magnetic flux profiles measured over the width of the channel. In Fig. \ref{fig:raw_dMdV} (a) we show a model magnetic flux profile obtained by convolving the SQUID point spread function with a boxcar function representing a uniform shift in the sample magnetization. In Fig. \ref{fig:raw_dMdV} (b-c), we show the raw magnetic flux data $\delta \Phi/\delta V_{TG}$ measured by scanning the SQUID over the channel while modulating $V_{TG}$ as described in the main text. In (b) we show $\delta \Phi/\delta V_{TG}$ acquired scanning on the line indicated in Fig. 2a. In (c) we show $\delta \Phi/\delta V_{TG}$ acquired in a different location separated by $\sim \SI{20}{\micro \meter}$. In black we show fits of the model flux profile in (a) to each of the line traces in (b) and (c) where the shape of the profile is the same for each fit and the amplitude of the shift in magnetization, $\delta M$ is allowed to vary. From $\delta M$ extracted from the fit, we calculate the response $\delta M/\delta V_{TG}$.

\end{document}